# Improving Radio Systems Efficiency via Employing SCRO-SOA (Squared Cosine Roll off Filter - Semiconductor Optical Amplifier Technology) in DWDM-RoF System

Bashar J. Hamza[1], Wasan Kadhim Saad[1], Mohamed Ahmed AbdulNabi[2], Waheb A. Jabbar[3], and Ibraheem Shayea[4]

[1]Al-Furat Al-Awsat Technical University, Engineering Technical College-Najaf, 31001 Najaf, Iraq.

[2]kufa university, Engineering -College-Department of Communication Engineering

[3]Faculty of Electrical & Electronics Engineering Technology, Universiti Malaysia Pahang, Pekan 26600, Malaysia.

[4]Electronics and Communication Engineering Department, Faculty of Electrical and Electronics Engineering, Istanbul Technical University (ITU), 34467 Istanbul, Turkey

E-mail: [coj.bash@atu.edu.iq , was.saad@atu.edu.iq , mohammeda.khaleel@uokufa.edu.iq , waheb@ieee.org , ibr.shayea@gmail.com]

## Abstract

The increasing demand on the internet trafficking to meet the demands of video streaming and mobile communication has exerted too much pressure. Accordingly, this demand will result in providing high bandwidth which in turn increase the use of cells on the network. However, the current networks do not meet the requirements of the necessary data rate. Therefore, the Dense Wavelength Division Multiplexing (DWDM) network and the Radio over Fiber (RoF) technology are the ideal solution for providing the necessary data rate needed in the current networks. The DWDM, will increase the transmission distance and will increase the data rate; nonetheless, the DWDM network will be compromised especially by the Non-linear effects. This study in intended to propose a system to find solutions for the issues of increasing the data rate and for reducing the nonlinear effects. There are number of technologies that could be adopted to fix such issues, which includes; Optical Phase Conjugation (OPC), Semiconductor Optical Amplifier (SOA), and Digital Signal Processing (DSP). Notwithstanding, DSP-128 QAM-SCRO (Squared Cosine Roll off) filter has been used in this study and then compared the results with Semiconductor Optical Amplifier (SOA) technology established which one of them is optimal to increase the data rate and to eliminate the nonlinear effects. The performance of the proposed system has been evaluated in terms of the Error Vector Magnitude (EVM), Bit Error Rate (BER),

Q Factor and Eye-Diagram. The EVM has been measured in proposed system prior to employing the DSP-128 QAM-SCRO technology, and it has reached the value 22.5%. However, when the EVM has been measured after employing the DSP-128 QAM-SCRO technology, it has reached the value 8.5 %. Similarly, the values of BER and Q factor have been found to be acceptable and in accordance with ITU-T. Finally, the proposed system has been simulated via the software Optisystem 17.

# I.Introduction

The accelerating growth in internet traffic currently is the ultimate challenge for radio access networks (RANs). The optical fiber technology has been widely spread in the last decades due to their ability for carrying high data rate for long distances with relatively low attenuation. Also, the optical fiber networks have various features in comparison to the traditional networks, RANs. The optical fiber networks are less affected by noise. They are not affected by the radio interference, and they are considered as totally isolated medium [1]. Several technologies have been developed to provide bandwidth for Tera per second in the optical fiber networks such the DWDM. Therefore, mobile network operators (MNOs) prefer to merge optical with radio technologies (Wi-Fi) to provide services for the end users with high data rate for long transmission distances [2]. And the combination of the Wi-Fi, the wireless, with optical fiber, wire, resulting the technology of the radio over fiber. It is considered as future solution for providing effective bandwidth in terms of cost and maintenance [3].

The DWDM of multiple density technology is used for the vert short distances between channels. Usually, their bandwidth is of 0.1- 0.8 Nanometer; therefore, DWDM has the possibility to connect tens of optical channels utilizing one fiber. This possibility increases the connectivity of the optical fibers [4]. The ease of using DWDM networks is due to the opportunity of utilizing very narrow spectrum known as the DFB. The DWDM has two features: first, the width of transmission spectrum, the mm-wave which facilitates channel distancing;

second, and the number of the wavelengths is significant which results in increasing the number of stations in the telecommunication networks [5]. Moreover, in the optical networks, there has to be a unique wavelength for each base station (BS) which can be retrieved and used in the main transmission [6][7]. As a result, it is possible to suggest an efficient and comprehensible DWDM-RoF structure especially in terms of optical fiber data rate, cost, and number of the wavelengths supporting the antennas distributed along the network [8].

Frequently, disconnections might occur during sending the signal mainly due to the nonlinear effects and dispersion. The nonlinear effects and dispersion lead the system to underperform where there the sent signal will deteriorate in the carrying channels resulting in certain interactions within the carrying medium which in turn responsible for producing the Kerr Effect. Mainly, the refractive index is responsible for generating the Kerr Effect. It is important to notice that the refractive index changes when the density of the optical power changes too [9].

The nonlinear effects occur if the light travel through the optical fibers causing the refractive index interact with medium of the optical fiber. The Kerr effect is most evident when the signal is modulated causing the phase to change because of the material sensitivity of which the optical fibers are made. The nonlinear sensitivity is greatest when there is signal interference and the signal is polarized, that and only then the optical pulse is generated. Once the optical pulse is generated, the end result would be one of the following effects, the SPM, XMP, or FWM [10][11].

The remaining part of the paper is arranged as follows: Section II introduces the related work. Section III presents description of the proposed system design. Section IV describes the simulation system and discussion of results. Section IV present the conclusion remarks of this paper.

II. BACKGROUND AND RELATED WORKS

Considerable research has been conducted investigating the topic at hand from different angles. For example, in [12], researchers extensively discussed one nonlinear effect, namely the FWD on a DWDM-RoF system of 32-channel 40 GBS. In this study, researchers analyzed the FWM effect on the inter-channel spacing, the level of input power, the area of the effective fiber as well as the types of modulation. It is crucial to point out that this study pioneered trials in minimizing the aforementioned effects. Therefore, this study suggested an optimization for setting up a DWDM system via employing the SPO technique.

The authors of [13] investigated possible schemes the enhancing wireless telecommunication systems of high data rates. For that purpose, the authors think that adopting a DWDM system would meet the demands of applications that require high data rate. In this paper, the researchers utilized certain amplifiers, EDFAs, and compensation technique namely the DCF in order to test the performance of the system when both the length of the optical fiber as well as the bit rates vary. The authors found out that the attenuation and dispersion are the most degrading effects for the system performance. Accordingly, the authors adopted the DCF technology for mitigating dispersion, and they utilized EDFAs to reduce attenuation and scattering.

In the study [14], the authors investigated the possibility of improving the efficiency of combining the wireless with wire networks, specifically the RoF. However, the challenge is how to assign wavelengths to light paths when designing a DWDM system. And to solve this problem, the authors suggest utilizing algorithm BCO-RWA with modifications since algorithm BCO-RWA does not tune in the FWM effect.

In a study conducted in [15], the researchers investigate the performance of DWDM system when using both two different modulation, namely the return-to-zero (RZ) as well as the non-return-to-zero (NRZ). Mainly, the researchers aimed at estimating and mitigating the FWM effect while adopting in-line optical phase conjugator (OPC). It is worthy to indicate that the FWM effect was estimated via employing actual fiber link which does not exhibit losses due to nonlinearity and attenuation. They found out that the power of the FWM effect could be suppressed when a destructive interference is introduced between the two halves of the OPC, the first half and the second one.

In a study carried out in [16], the researchers investigated the FWM effect on the performance of a DWDM-RoF system. They adopted ODSB, Optical Double Side Band, along with Mach–Zehnder modulator of dual drive for both external and direct modulations. Additionally, the authors presented an analysis for the performance of the proposed system in the presence of the FWM effect for various parameters including channel spacing, data rate, input power, optical amplifier gain, and finally fiber length.
.

In a different study, [17], researchers proposed a DWDM system of 100 and 200 Gbs while introducing (PM-QPSK), Polarization-multiplexing quadrature phase shift keying, as a key component. Such system performs variously depending on the pulse shaping technologies such as the RZ, NRZ, or CSRZ (Carrier Suppressed Return-to-Zero. The researchers analyzed the extent to which the PM-QPSK is tolerant to both of the linear as well as the nonlinear effects such as ASE noise, distortion resulting from optical filtering, crosstalk, CD, and PMD in addition to the Kerr optical impairments.

In accordance with previous studies, the authors of [18] hypothesized that the FWM and XPM are among the most critical effects impacting the performance of a (DWDM) system. It has been established that the FWM results in a producing a new wave which is responsible for wasting power, and it could lead to crosstalk. Also, power variation might result in the XPM effect which, in turn, is responsible for pulse distortion and hindrance in transmission capacity. Thus, the main concern of this study was to show the XPM and FWM effects on variety of dispersion ranges in RZ and NRZ signals. .

In a more recent study, authors of [19] investigated the possibility of meeting the demands of high input power and increasing the number of channels of low spacing in the Next Generation Optical Communication Networks. And to meet these demands, the author suggested utilizing a DWDM system. However, reducing channel spacing and treating nonlinear effects, namely the FWM, was the main challenge for the researchers. Accordingly, they suggested various modulations to solve the problem mentioned above. And the authors evaluated their proposed system in terms f the Q-Factor and the Optical Signal-to-Noise Ratio.

Furthermore, authors of [20] conducted a study aiming at enhancing hybrid optical amplifiers, namely the SOA, EDFA, and RAMAN to operate in a DWDM system. This system also operates at the extreme level of L and U wavelengths. They utilized a booster-an optical amplifier which a semiconductor- in conjunction with the EDFA and RAMAN amplifiers in order to enhance the performance of the system. In their proposed design, 40 channels at 10 Gbps within the frequency range of the hybrid amplifier was characterized in accordance with the maximum gain, gain flatness, NF standing for the amplifier noise figure. Finally, the authors compared forward and backward types of pumping for the RAMAN amplifier when various lengths of RAMAN fiber were used.

In the most state-of-the-art study, the authors of [21] worked on utilizing Polarization Mode Dispersion (PMD) via employing PMD emulator at the input in order to suppress FWM effect in a DWDM system. This study was conducted to take advantage of the high performance of data networking as well as the security of DWDM technologies. However, the demands of high input power in addition to the low channels spacing can result in nonlinear effects such as the WDM which is the main reason in generating the FWM effect.

This study aims at enhancing the performance of the radio systems over the optical fiber when combined with the DWDM in presence of the nonlinear effects. It also aims at compensating for these nonlinear effects as well as fixing them via adopting the DSP-128 QAM-SCRO (Squared Cosine Roll off) technology. Moreover, this paper tries to establish the extent to which increasing the data rate and the transmission distance could be increased. Generally, in the proposed system, the main goal is to find a replacement for the amplifiers, repeaters, and compensators which are mathematically complicated such the OPC. Such amplifiers, including the SAO, also known for their high losses. In other words, the aim is to provide new filtering methods which are compatible with telecommunication systems.

III - Description of the Proposed System DWDM-RoF

Figure (1) illustrates the structure of the DWDM-PON-RoF network for the propose system. It is an optical system which consists of three parts: the Optical Line Network (OLT), Optical Network Unit (ONU), and Optical Distribution Network (ODN). The linear and nonlinear distortions will be exerted on the system such as attenuation, polarization mode dispersion (PMD), and dispersion as linear effects in addition to the nonlinear ones such as the SPM, XPM, and FWM.

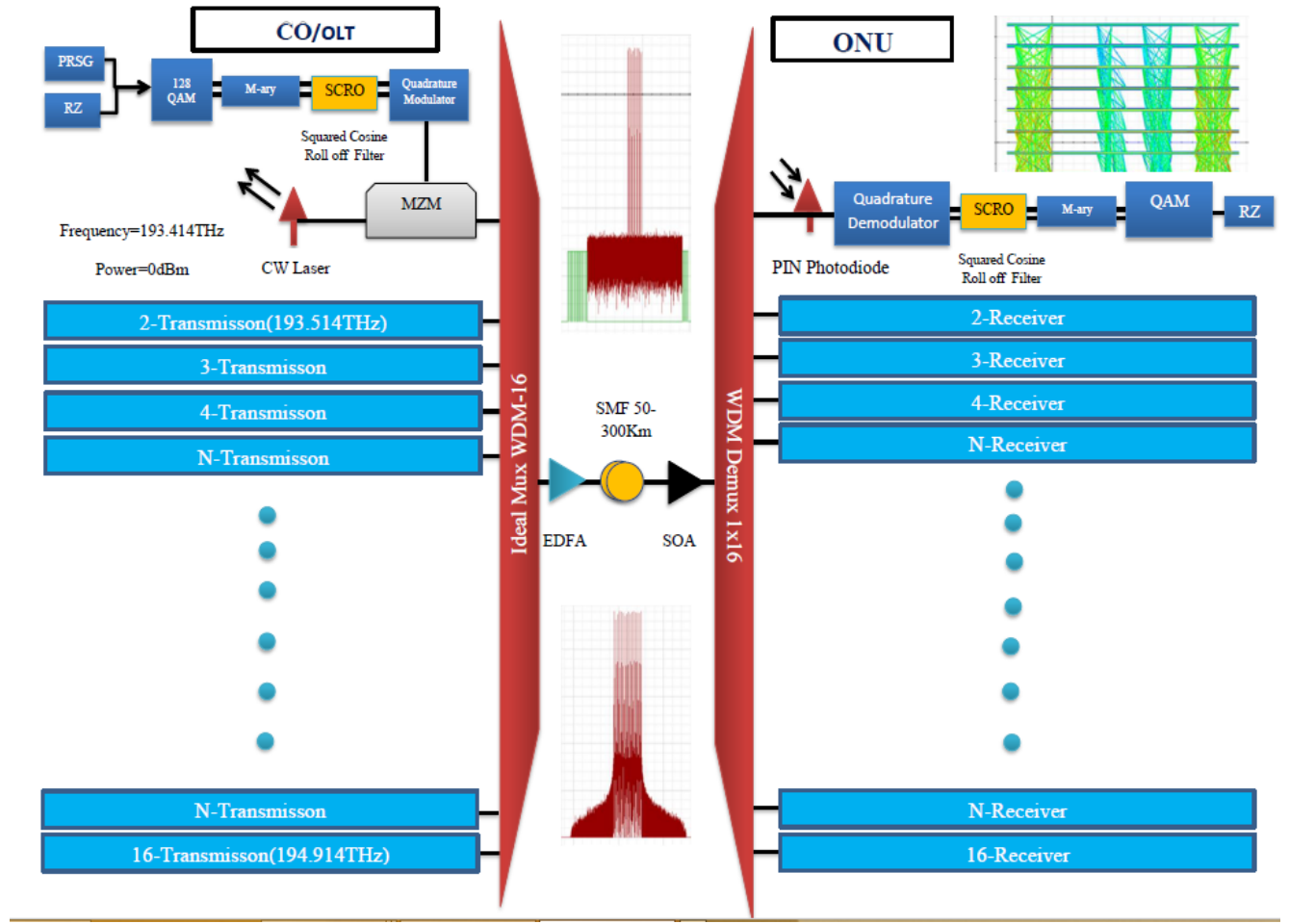

Figure (1) The proposed DWDM- RoF architecture system

The proposed system operates in accordance with the digital signal processing (DSP) technology. The OLT contains the Transmission X (TX) as in figure (2), which is responsible for sending signals of fixed power for testing the proposed system. It is worth noting that the TX contains 16 channels and 16 wavelengths which range from 193. 414 THz to 194.914 THz, where the differences between any two wavelengths is less than or equal to 100 GHz and this was intended to insure generating the DWDM and to ensure inter-channel distancing. Additionally, 16 Pseudo-Random Bit Sequence Generators (PRBSG) have been added to the TX. The PRBSGs are responsible for generating random encrypted data without the pulse generator return zero (RZ). In this proposed system, each one of the 16 channels have been modulated from the electric form to the optical one. Therefore, 16 electric modulators of the 128 QAM type have been utilized where the output represented as (I). The (I) is connected with Quadrature Amplitude Modulation (QAM) to generate a series of M-ray connected to the DSP-128 QAM-SCRO (Squared Cosine Roll off) filter, which is, in turn, connected to16 Quadrature Modulators. Eventually, all the aforementioned elements of the TX circuit are connected to both of the Mach-Zehnder Modulator and the CW laser.

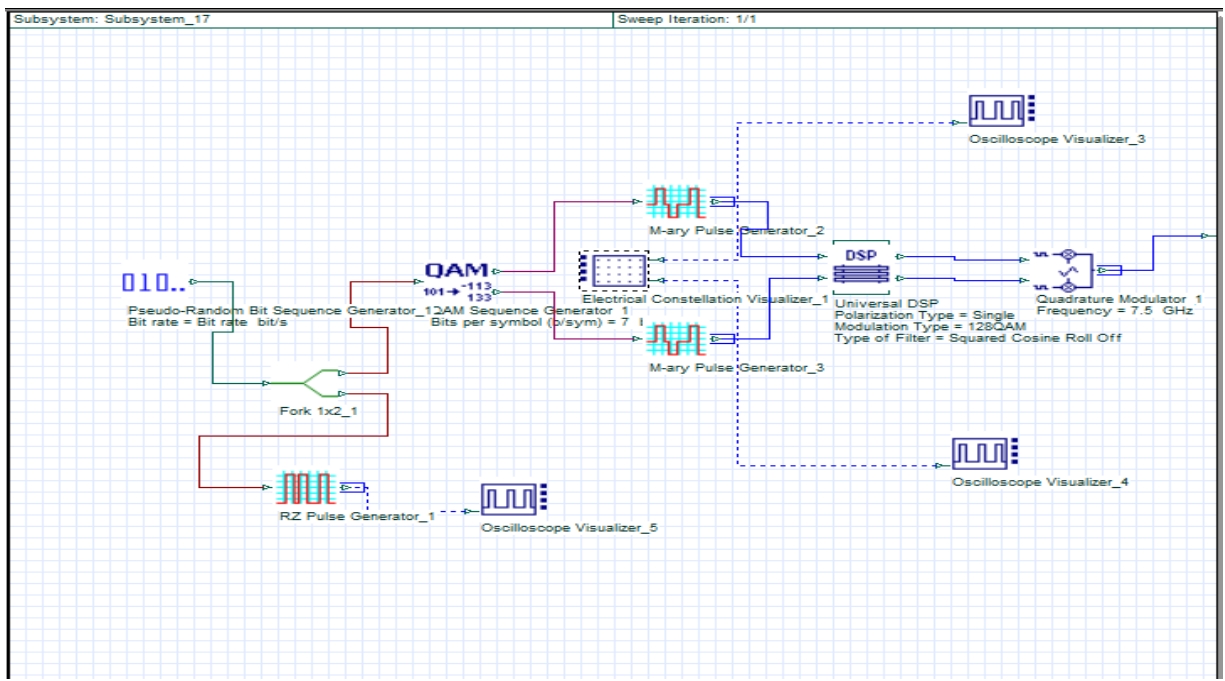

Figure (2) Transmission (OLT)

The second part is the Optical Distribution Network. It contains only two components: the first one is the carrying medium that is the optical fiber which ranges from 50-300 Km in length. The second component is the Er-doped fiber amplifiers by considering numerical solutions (EDFA). As Illustrated in figure (3).

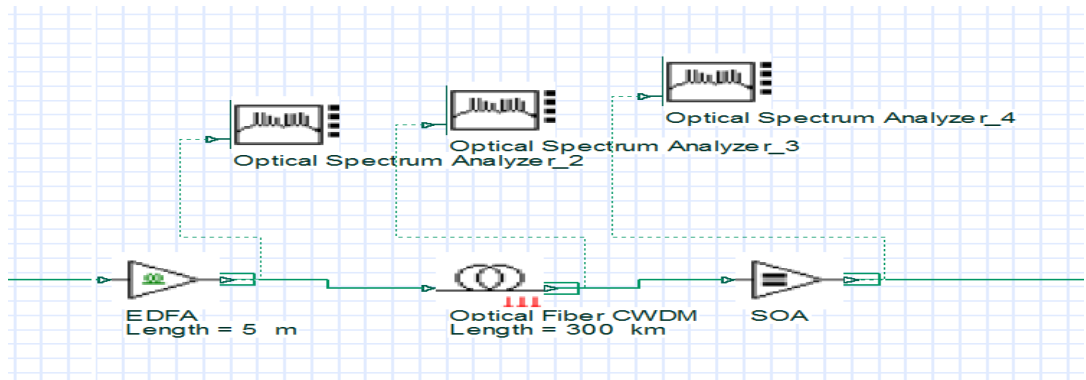

Figure (3) Optical Distributed Network (ODN)

The third part is the Optical Network Unit (ONU). It contains 16 PIN Photodiode which lessens light density in long range networks. The ONU also contains 16 Quadrature Demodulator in addition to 16 DSP-128 QAM-SCRO (Squared Cosine Roll off) filter. The ONU also contains decoding devices for the QAM, the M-ray detectors, pulse generator RZ, and finally, the Quadrature Demodulator as illustrated in figure (4).

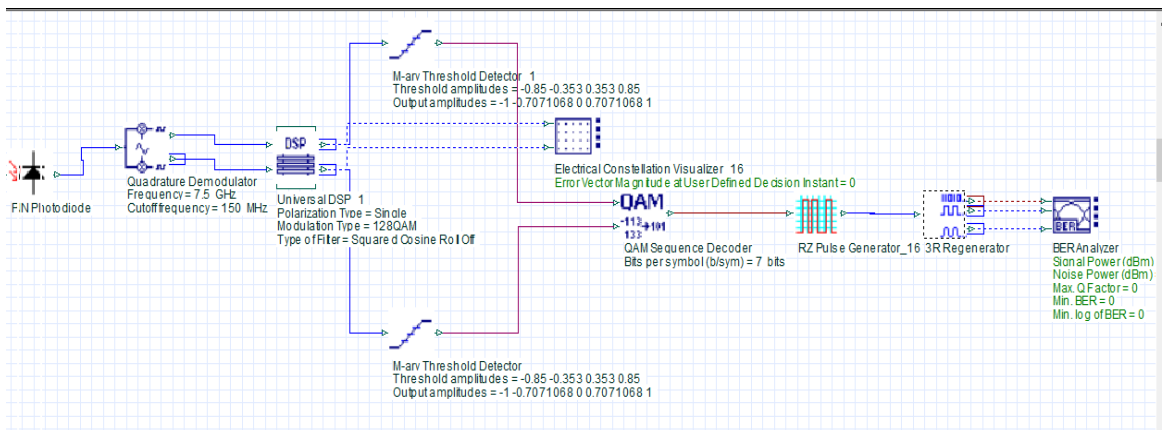

Figure (4) Receiver (ONU)

## IV- Results and Discussion

To achieve the aims of this investigation, the performance of the DWDM-RoF has been simulated twice. In both of two simulations, the performance of the systems has been test under the nonlinear effects. The first simulation has been carried out adopting the SOA technology to reduce the nonlinear effects. The second simulation has been conducted utilizing DSP-128 QAM-SCRO (Squared Cosine Roll off) filter for the purpose of reducing the nonlinear effects. The main objective of these two simulation was to discover what might happen to the light spectrum when it traveled for long distances. For that purpose, the density of wavelength has been measured through transmission distance from 50-300 Km, where the wavelengths are uplinked with 300 Gpbs under presence the nonlinear effects as illustrated in figure (5).

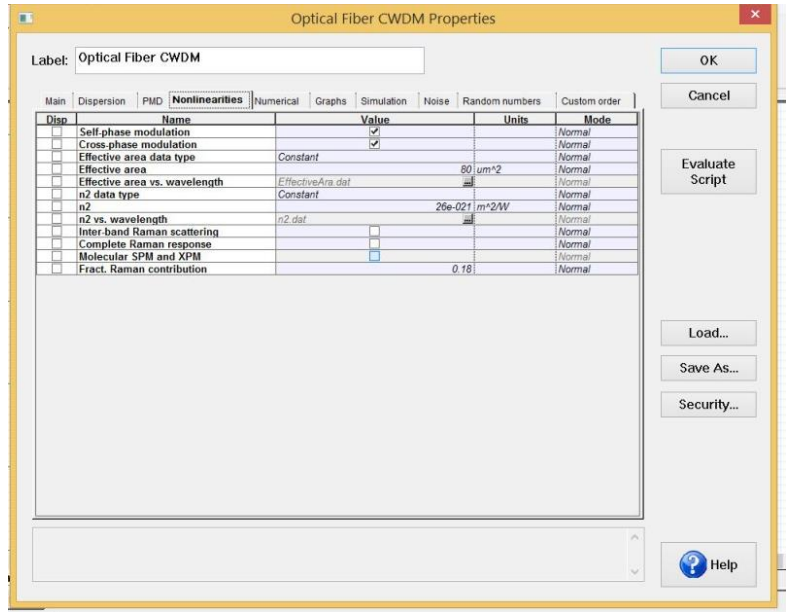

Figure (5) SPM, XPM , FWM parameters

The form of the light spectrum is shown in figure (6) when using the SOA technology due to transmitting wavelengths that range from 193.414 THz to 194.914 THz.

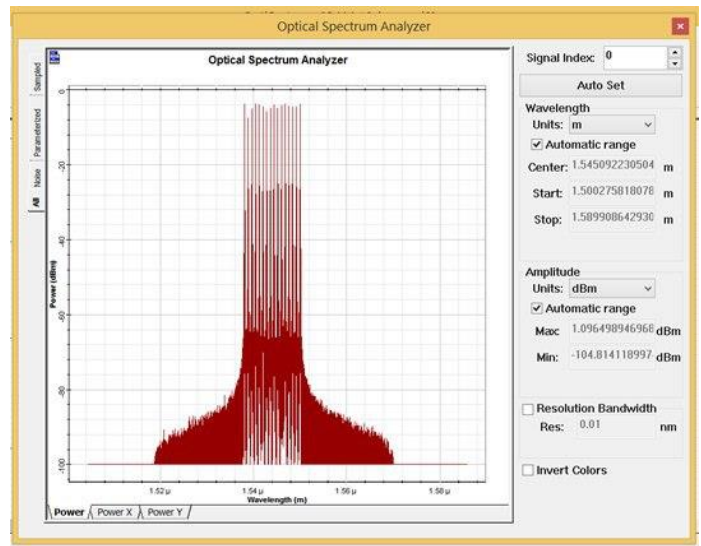

Figure (6) Input signals frequency spectrum

Transmission thorough the optical fiber has been evaluated in varying distances that range from 50-300 Km. And after checking the nonlinear effects, the SPM, XPM, and FWM in the optical system DWDM-RoF in equally-distanced channels. Accordingly, the nonlinear effects will take the form of the second order harmonic at the output signals. Due to the nonlinear effects, the second order harmonic will become of considerable importance in accord with the transmission distances where the form of the spectrum frequency will be as shown figures (7a, 7b, and 7c). Figure (7d) shows that the second order harmonic will have greatest impact on the performance of the system.

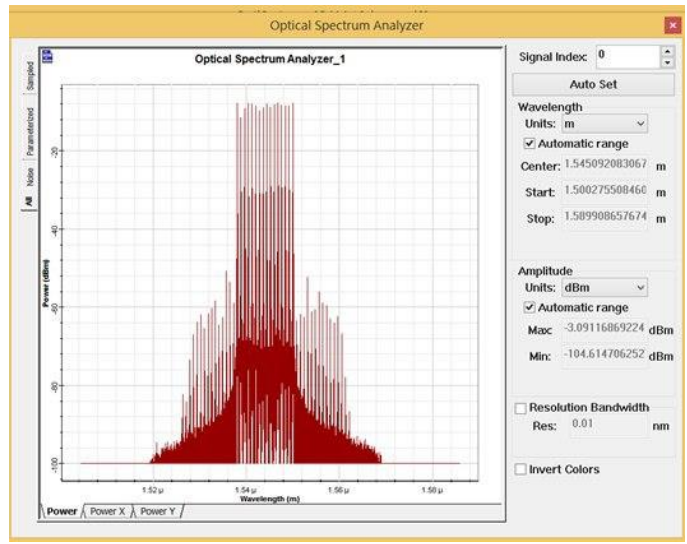

Figure 7a- Output signal frequency spectrum

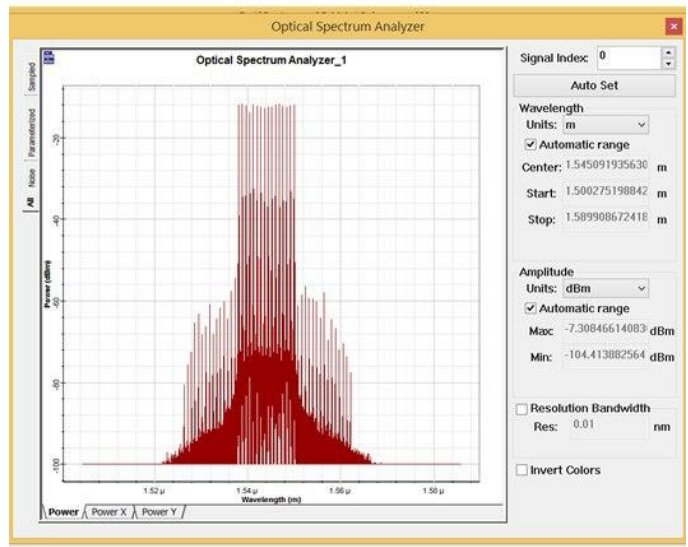

Figure 7b Output signal frequency spectrum

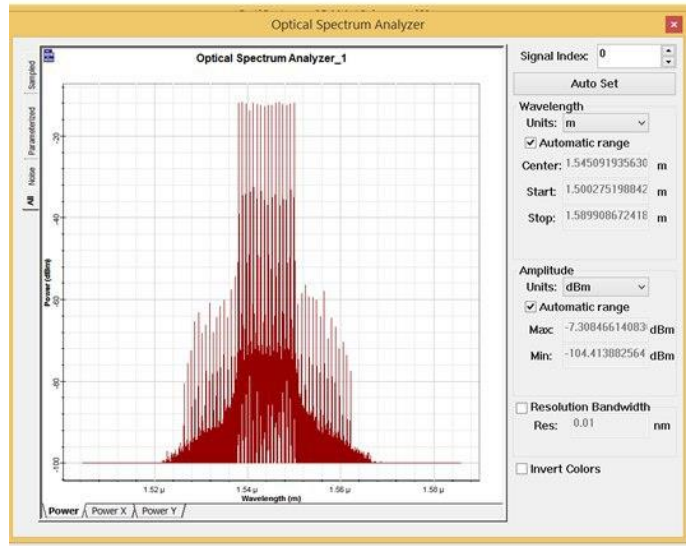

Figure 7c Output signal frequency spectrum

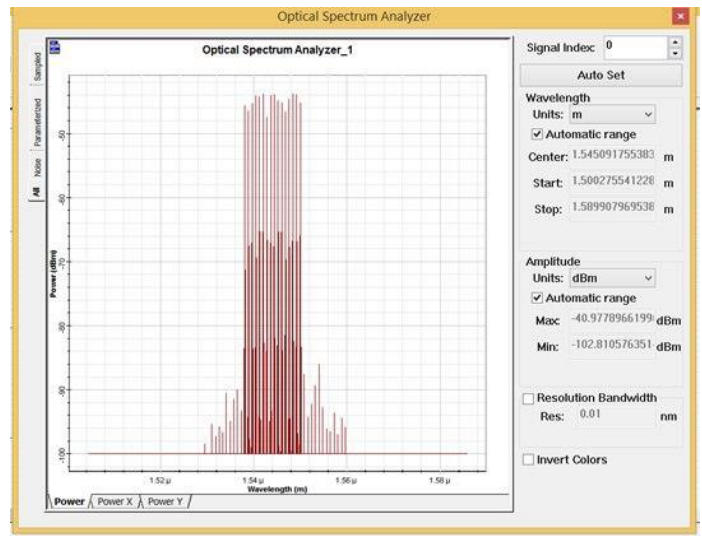

Figure 7d Output signal frequency spectrum

The quality of the optical signal has been also evaluated in the proposed system under the presence of the nonlinear effects listed above via using graphical representation, the Eye-Diagram at 300 Km as illustrated in figure (8).

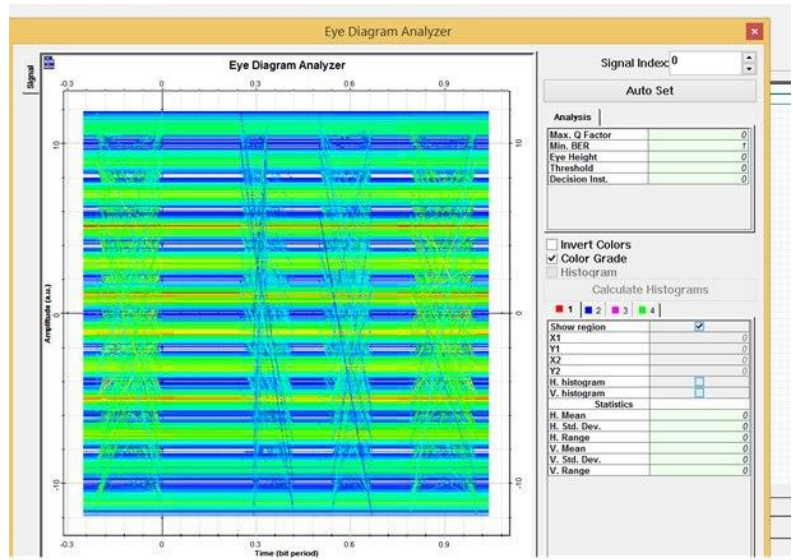

Figure (8) Eye-Diagram in presence of the nonlinear effects

After observing how the nonlinear effects might impact the performance of the system while utilizing the SOA technology, the results above show that the overall quality of the system deteriorates. The results also show that the signal fades out clearly due to the second order harmonic; therefore, an alternative technology, the DSP-128 QAM-SCRO (Squared Cosine Roll off) filter, will be utilized which operates according to the principles of digital signal processing for the same data rate at long distances without using the compensation technology, the SOA. A comparison has been made between the results of the DSP-128 QAM-SCRO (Squared Cosine Roll off) filter and the SOA technology.

When utilizing DSP-128 QAM-SCRO (Squared Cosine Roll off) filter, the shape of the light spectrum for both of the input and output will be as illustrated in Figure (9). Accordingly, the second order harmonic has been reduced considerably as shown in figure (10a, 10b, 10c, and 10d).

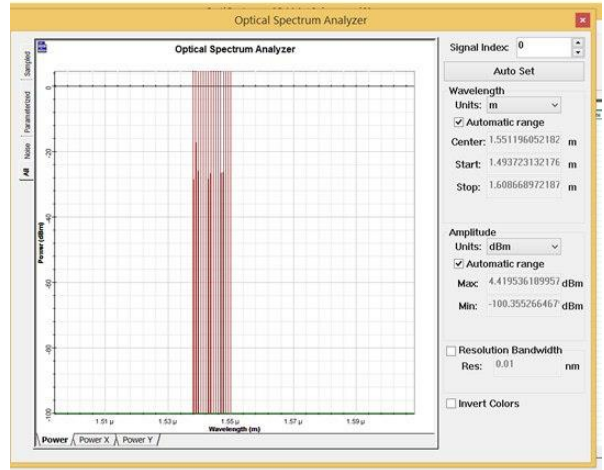

Figure (9) Input signals frequency spectrum with QAM-SCRO

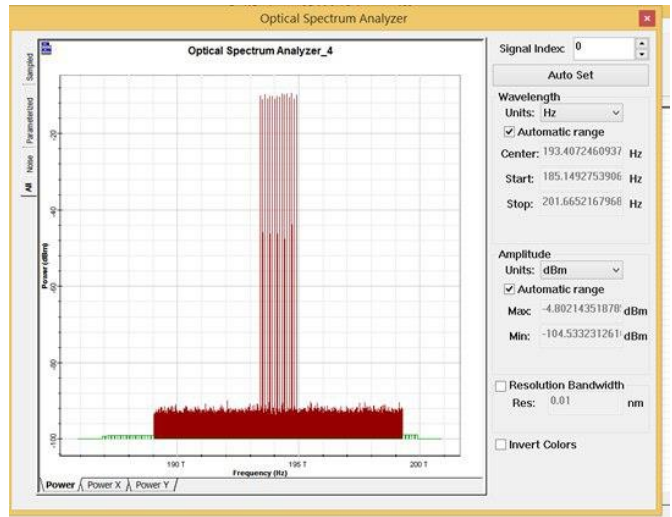

Figure 10a Output signal frequency spectrum

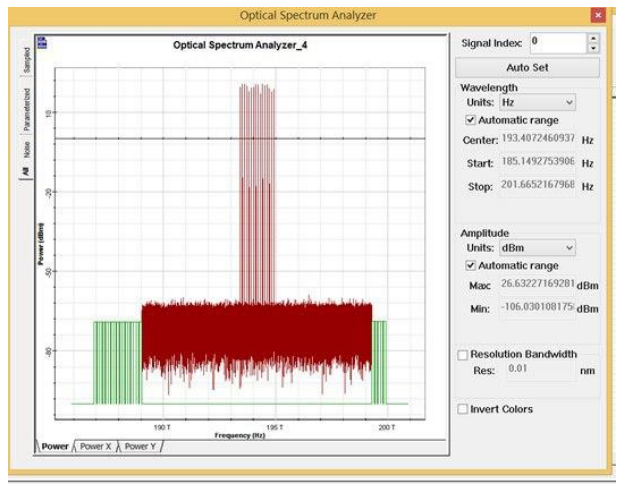

Figure (10b) Output signal frequency spectrum

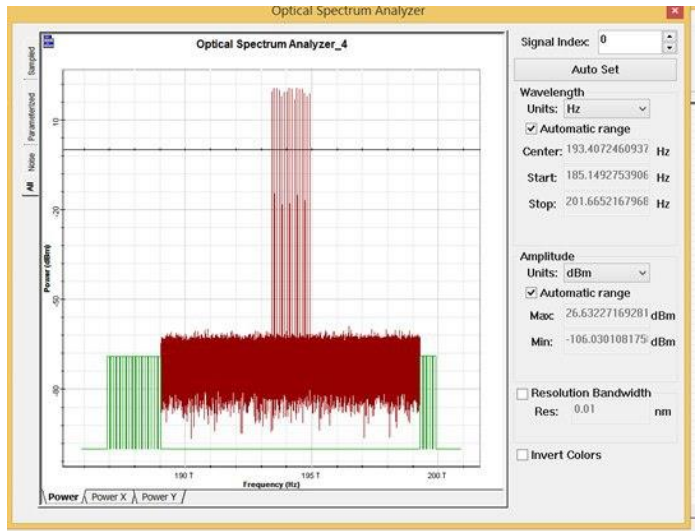

Figure 10c Output signal frequency spectrum

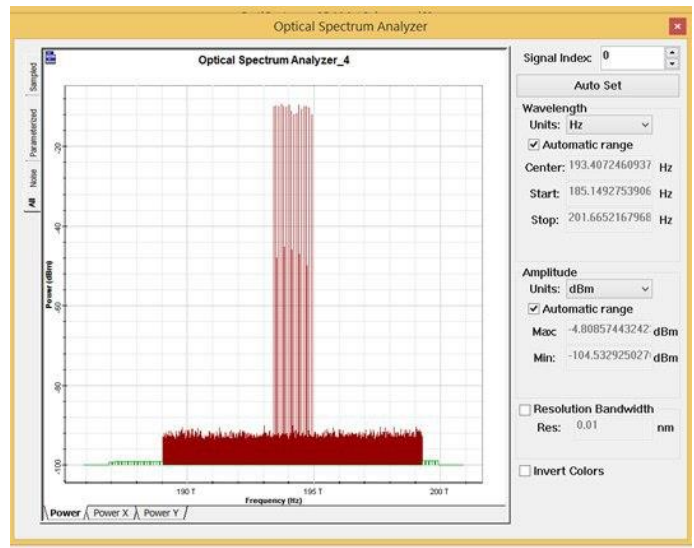

Figure 10d Output signal frequency spectrum

Additionally, while utilizing the DSP-128 QAM-SCRO (Squared Cosine Roll off) filter The quality of the optical signal has been also evaluated in the proposed system under the presence of the nonlinear effects listed above via using graphical representation, the Eye-Diagram at 300 Km as illustrated in figure (11).

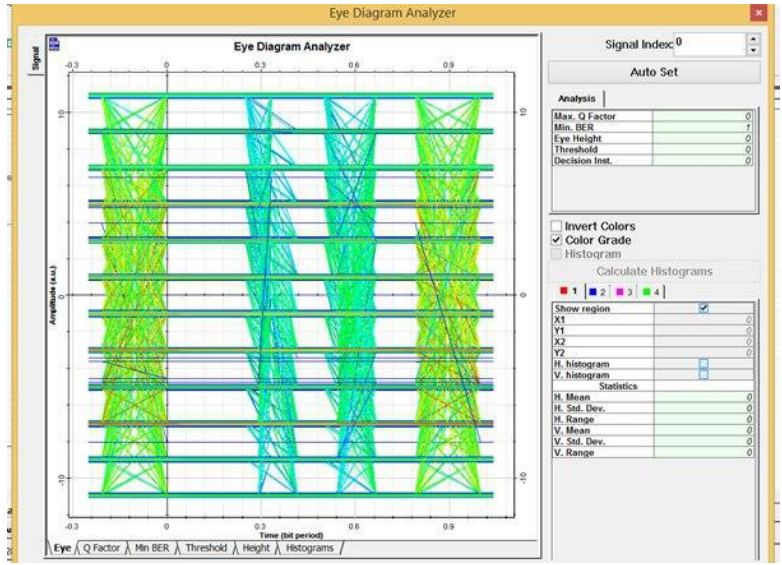

Figure (11) Eye-Diagram in presence of the nonlinear effects

Figure (12) shows the relationship between the BER and the wavelength for how the nonlinear effects impact the performance of the system when each nonlinearity reduction technologies. It is evident that the SCOR filter aids in achieving acceptable BER value which is equal to 836E-13 at wavelength of 193.414; however, the when utilizing the SAO technology for the same wavelength, the BER is equal to 178E-7 as illustrated in figure (12).

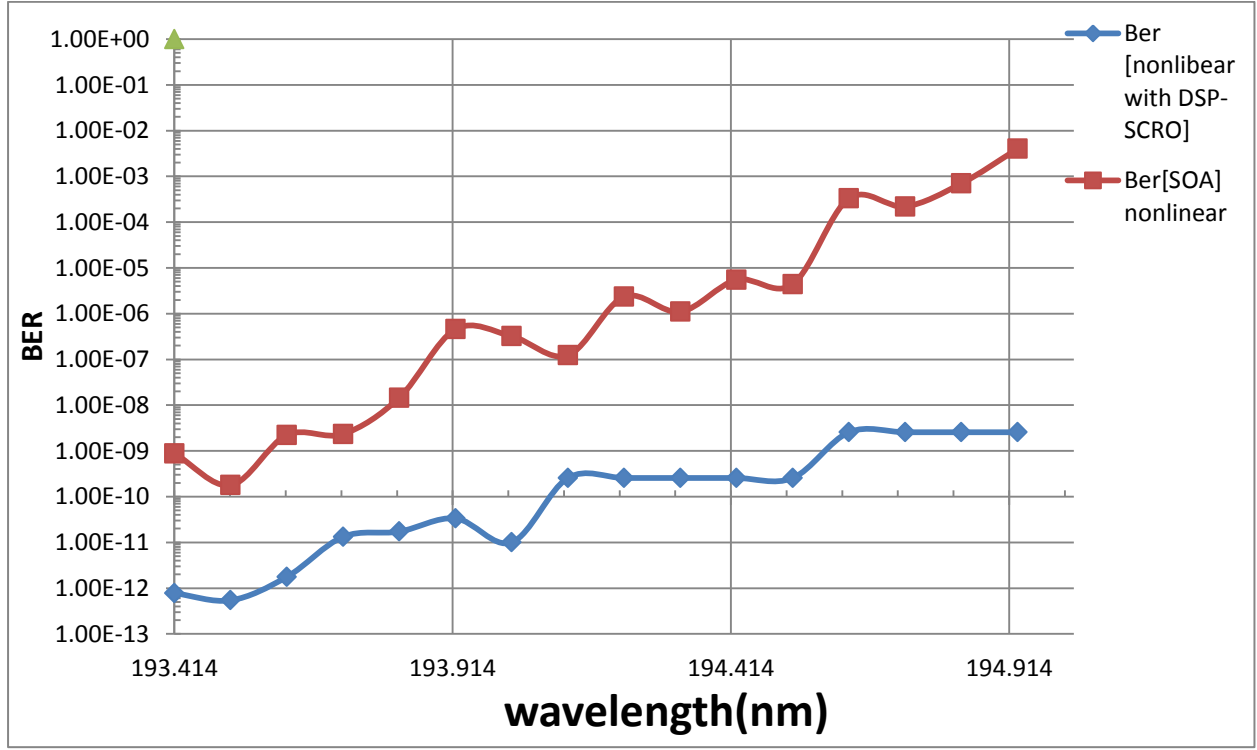

Figure (12) BER to Wavelength in the presence of the nonlinear effects

.

When measuring the Q factor and the BER parameters in relation to distance, it is evident that the SOA technology does not operate at the distance of 200Km where the values of the BER and Q factor are 2.34E-7 and 5.44 respectively. Nonetheless, when utilizing the SCRO, the system operates properly coupled with significant reduction in the nonlinear effects where the values of the BER and Q factor are 10.40 and 1.32E-10 respectively as illustrated in figures (13 and 14).

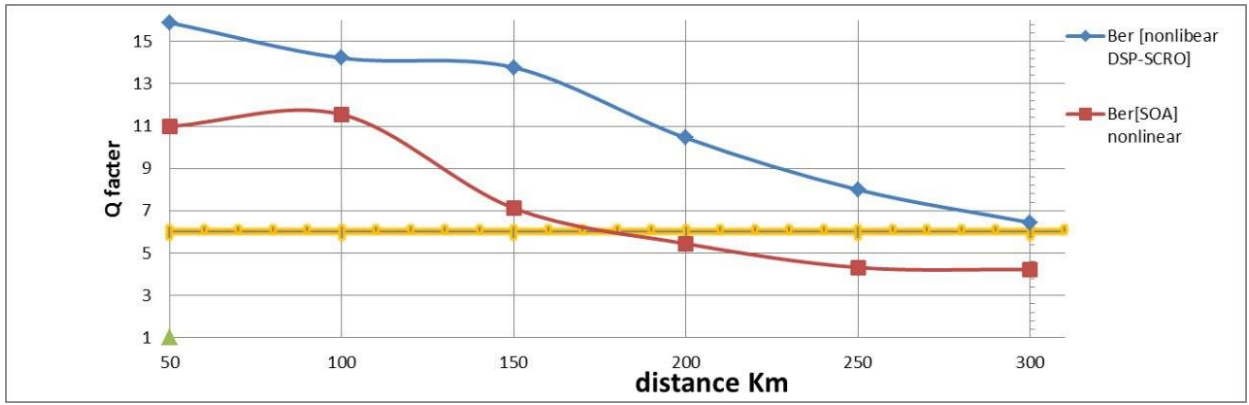

Figure (13) Q Factor to Distance in the presence of the nonlinear effects

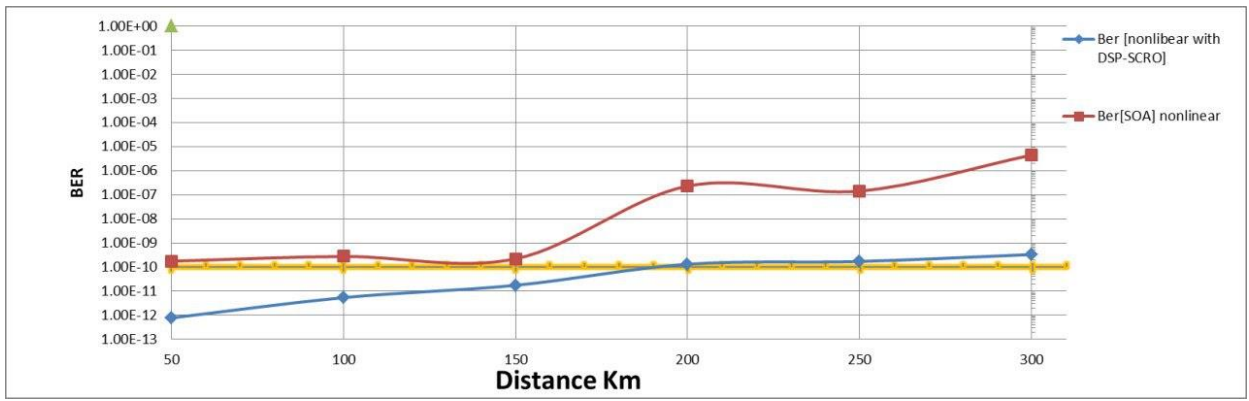

Figure (14) BER to Distance in the presence of the nonlinear effects

As shown in Figure (15), the EVM value is 22.5 when using the SAO technology at the distance of 300 Km. This value indicates the system does not operate properly, and the system exhibits losses in every stage, i.e. transmission, reception, and the carrying medium.

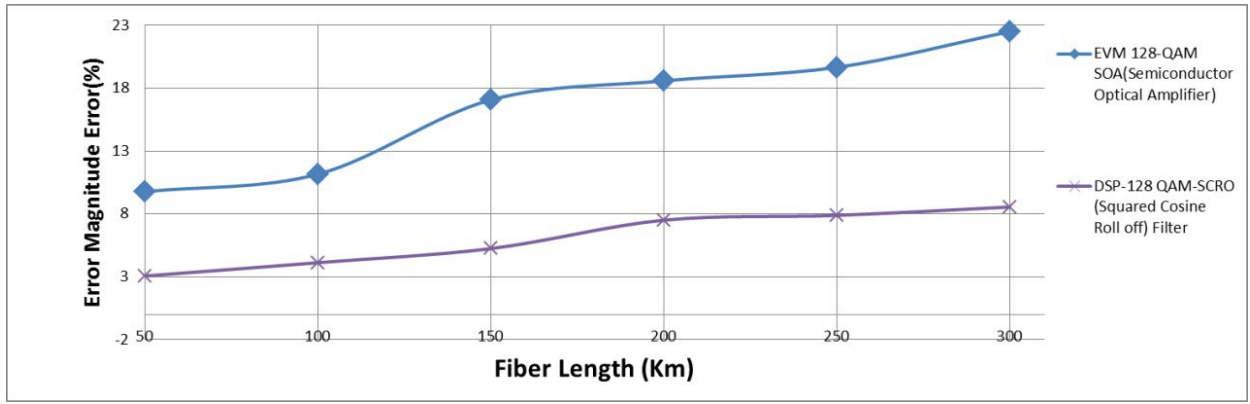

Figure (15) comparison of EVM to Distance

## V- Conclusion

To sum up the discussion presented above, the following remarks can be drawn. In the foreground, the DSP-128 QAM-SCRO (Squared Cosine Roll off) Filter is considered one of the best technologies which has been tested in the optical communication systems for the DWDM-PON-RoF. It is considered the best technology for achieving the optimal values for both of the Q factor and the BER at the distance of 300 Km. Further, when employing the DSP-128 QAM-SCRO (Squared Cosine Roll off) in the proposed system, the performance of the system has enhanced significantly by measuring the EVM in the system where it was reduced from 22.5 to 8.5. Conversely, when using the SOA technology, the EVM was never reduced at the distance of 300 Km. Furthermore, the DSP-128 QAM-SCRO (Squared Cosine Roll off) Filter aids in mitigating the nonlinear effects, namely the SPM, XPM, and FWM, without having to using the modulations and complications found in the other compensation technologies which are known for their high cost and their complicated maintenance.

At the basic level, the investigation is successful in terms of two considerations. The first consideration is that the second order harmonic has been totally eliminated. The second one is that the DSP-128 QAM-SCRO (Squared Cosine Roll off) filter does not allow the inter-symbol interference due to the fact that it possesses unmatched speed in avoiding the delay resulting from the interaction between the light and the carrying medium.

Finally, for the future work, it is safe to suggest investigating the performance of the DWDM-PON-RoF system when the DSP is 256 QAM and 512 QAM in order to increase the transmission distance for more than 500 Km with data rate 500 Gpbs while employing multi-usage filter specifically the Cosin RoLL Off.